\documentclass[conference]{IEEEtran}
\IEEEoverridecommandlockouts
\usepackage{cite}
\usepackage{amsmath,amssymb,amsfonts}
\usepackage{algorithmic}
\usepackage{graphicx}
\usepackage{textcomp}
\usepackage{xcolor}
\usepackage[acronym]{glossaries}
\usepackage{balance}
\newacronym{pr}{PR}{Probe Request}
\newacronym{prb}{PRB}{Probe Requests Burst}
\newacronym{mac}{MAC}{Media Access Control}
\newacronym{ie}{IE}{Information Elements}
\newacronym{pnl}{PNL}{Preferred Network List}
\newacronym{ifat}{IFAT}{Inter-Frame Arrival Time}

\newacronym{os}{OS}{Operating System}

\newacronym{ml}{ML}{Machine Learning}
\newacronym{tpr}{TPR}{True Positive Rate}
\newacronym{fpr}{FPR}{False Positive Rate}
\newacronym{auc}{AUC}{Area Under the Curve}
\newacronym{roc}{ROC}{Receiver Operating Characteristic}
\newacronym{rmse}{RMSE}{Root Mean Square Error}
\newacronym{pb}{PB}{Pairwise Boosting}
\newacronym{apb}{APB}{Asymmetric Pairwise Boosting}
\newacronym{dbscan}{DBSCAN}{Density-Based Spatial Clustering of Applications with Noise}
\newacronym{hdbscan}{HDBSCAN}{Hierarchical DBSCAN}
\newacronym{optics}{OPTICS}{Ordering Points To Identify the Clustering Structure}

\usepackage{lipsum}

\def\BibTeX{{\rm B\kern-.05em{\sc i\kern-.025em b}\kern-.08em
    T\kern-.1667em\lower.7ex\hbox{E}\kern-.125emX}}

\usepackage[hidelinks]{hyperref}

\usepackage{todonotes}

\usepackage{fontawesome5}

\usepackage[ruled,lined,linesnumbered]{algorithm2e}

\SetCommentSty{it}
\begin{document}

% \title{A Compact Fingerprint for Wi-Fi Probe Requests Based on Asymmetric Pairwise Boosting}

\title{Compact Probe Request Fingerprinting with Asymmetric Pairwise Boosting
}

\author{\IEEEauthorblockN{Giovanni Baccichet, Fabio Palmese, Alessandro E. C. Redondi, Matteo Cesana}
\IEEEauthorblockA{\textit{Dipartimento di Elettronica, Informazione e Bioingegneria, Politecnico di Milano} \\
email: \{name.surname\}@polimi.it}
}

\maketitle

%vital for optimizing infrastructure, enhancing security, and improving public services.

\begin{abstract}
Probe Requests are Wi-Fi management frames periodically sent by devices during network discovery. %These frames are crucial for applications in urban planning, human mobility analysis, and retail analytics. 
Tracking Probe Requests over time offers insights into movement patterns, traffic flows, and behavior trends, which are keys in applications such as urban planning, human mobility analysis, and retail analytics. To protect user privacy, techniques such as MAC address randomization are employed, periodically altering device MAC addresses to limit tracking. However, research has shown that these privacy measures can be circumvented. By analyzing the Information Elements (IE) within the Probe Request body, it is possible to fingerprint devices and track users over time.
This paper presents a machine learning-based approach for fingerprinting Wi-Fi Probe Requests in a compact fashion. We utilize Asymmetric Pairwise Boosting to learn discriminating filters which are then used to process specific bit sequences in Probe Request frames, and quantize the results into a compact binary format. Extensive evaluation on public datasets demonstrates a two-order-of-magnitude storage reduction compared to existing methods while maintaining robust fingerprinting performance. %This advancement empowers researchers and practitioners to effectively analyze large-scale Probe Request datasets, unlocking valuable insights even in the face of privacy-enhancing measures. %Extensive experiments over public datasets of Probe Request frames demonstrate that the proposed approach reduces storage requirements by two orders of magnitude compared to the state of the art, while maintaining tracking accuracy with minimal loss. By effectively balancing the storage-accuracy requirements, our method enhances the management of extensive Probe Request datasets, making them more useful for long-term analysis and practical applications, even in the presence of advanced privacy measures. \\
\end{abstract}


\section{Introduction}
\label{sec:introduction}

%With billions of Wi-Fi-enabled devices shipped annually, we are living in a \textit{Global Village} where individuals are increasingly connected. 
Wi-Fi devices periodically transmit unencrypted frames, known as \gls{pr}, to search for available networks in their proximity. The collection of such frames enable entities such as urban planners, transportation authorities, and commercial enterprises to monitor and analyze urban flows, occupancy estimation, and extract useful metrics for resource planning \cite{Redondi2018}.

Extracting information from collected \gls{pr} frames is however very challenging nowadays, due to two main reasons:
\begin{enumerate}
\item \textbf{Storage Footprint}: The massive proliferation of Wi-Fi devices has led to an exponential growth in \glspl{pr} volume, overwhelming traditional storage and management capabilities. This growth in data presents significant challenges in terms of cost and complexity.
\item \textbf{MAC Address Randomization}: To avoid unauthorized eavesdroppers from tracking a device, one of the most popular countermeasures implemented nowadays is \gls{mac} Address randomization, where a device generates randomized addresses used in a pseudo-random pattern during the \gls{pr} emission process, hindering tracking.
\end{enumerate}
Methods to counteract \gls{mac} address randomization have been a focal point of recent research efforts \cite{di2016mind, tan2021efficient}: such methods attempt to extract fingerprints from \glspl{ie} contained in the payload of \gls{pr} frames, that can be later used to track a device without relying on its MAC address. However, little attention has been given to the problem of handling or compressing large volumes of \glspl{pr} efficiently. Compression techniques tailored to \glspl{pr} allow to significantly reduce the storage footprint of these requests without losing critical information. This not only helps in lowering storage costs but can also enhance data processing efficiency, enabling quicker and less consuming analysis. Ultimately, the ability to compress \glspl{pr} ensures that the benefits of monitoring and analyzing Wi-Fi signals can be sustained as the number of connected devices continues to rise.

Our goal within this study is to design a compact yet discriminative \gls{pr} fingerprint. In order to do this, we rely and adapt a \gls{ml} approach named \gls{apb}, initially used successfully for audio and video fingerprints \cite{Jang2009}.
\gls{apb} learns (i) discriminative portions of \gls{pr} as well as (ii) thresholds to quantize each portions to a 1-bit element: the concatenation of such bits forms a compact fingerprint which can be efficiently stored and matched for fast \gls{pr} processing.

In details, the main contributions of this work can be summarized as follows:

\begin{itemize}
\item We introduce a \gls{pr} binary fingerprint based on %bit-wise selection method using 
the \gls{apb} algorithm. To ensure reproducible research, our implementation is made publicly available\footnote{Code Repository: \href{https://github.com/GiovanniBaccichet/CompactProbes}{github.com/GiovanniBaccichet/CompactProbes}}. Moreover, the data used for training the proposed fingerprinting technique come from open datasets, detailed in \cite{Pintor2022_Dataset} and \cite{Baccichet2024}.
\item The proposed fingerprinting method requires less than $1\%$ of the memory used by state-of-the-art methods, with comparable fingerprinting accuracy.
\item We evaluate the performance of our methodology against leading \gls{pr} fingerprinting techniques. Our focus includes clustering metrics and memory requirements.
\end{itemize}

%These contributions not only provide a robust framework for improving data storage and analysis efficiency but also pave the way for more scalable and privacy-conscious applications in urban planning, transportation, and commercial analytics.

The rest of the paper is organized as follows: Section \ref{sec:related-work} discusses prior research on \gls{pr} and \gls{mac} de-randomization, as well as the \gls{apb} algorithm to create compact fingerprints. Section \ref{sec:methodology-fingerprint} outlines the proposed fingerprinting method, the proposed filters, as well as the fingerprint-matching pipeline. Section \ref{sec:dataset} details the binary data pre-processing. Section \ref{sec:experimental-results} discusses experimental results and presents a comparison with state-of-the-art techniques, while Section \ref{sec:conclusion} concludes the paper with final remarks and future research directions.

\section{Related Work}
\label{sec:related-work}

The study of \gls{pr} frames in Wi-Fi networks has become a well-investigated research area due to the significant privacy and security implications associated with these frames. \glspl{pr}, used to discover available networks in the transmission range of the devices, contain \gls{mac} addresses that can be potentially used to track devices. To address these privacy concerns, manufacturers and \gls{os} developers have implemented several countermeasures. These include randomizing the source address, minimizing metadata in \gls{pr} frames (e.g., obscuring the \gls{pnl} or eliminating non-essential fields), and periodically rotating frame content between probe bursts. This developments have sparked active research into \glspl{pr} fingerprinting methodologies for applications such as network optimization, security monitoring, and user analytics.
Early work by Vanhoef et al. \cite{Vanhoef2016} explored the use of temporal and spatial patterns in \gls{pr} frames. They identified that, despite the MAC randomization, devices often exhibit unique patterns in terms of SSID lists, inter-packet timing, and sequence numbers, which can be leveraged to link multiple \gls{pr} to the same device. Similarly, Matte et al. \cite{Matte2016} proposed a novel attack method that leveraged timing observations of \gls{pr} frames. They developed an \gls{ifat} signature to distinguish devices based on the rate at which they sent these requests. Recent research by Pintor et al. \cite{Pintor2022} provided a comprehensive analysis of \gls{ie} fingerprinting using \gls{dbscan} clustering, yielding promising results in a controlled laboratory setting. However, it highlighted a significant limitation: the misclassification of devices that share the same device model due to nearly identical IE fields. %This is due to the fact that IE encoding is performed by summing ASCII values of the selected \gls{ie}. This method can be problematic, as it frequently produces ambiguous results that hinder device fingerprinting. As an example, in the Vendor
%Specific Tags field, the HEX values \texttt{0050f208002800} (associated with probes labeled OppoFindX3Neo\_A), \texttt{0050f208006400} (associated with probes labeled XiaomiRedmi4\_B), and \texttt{0050f208009100} (associated with probes labeled XiaomiRedmi5\_J) all result in the same sum of $751$. This issue highlights the need for more sophisticated feature extraction and hashing techniques that can differentiate between features more effectively, thereby reducing the likelihood of such collisions and improving the precision of device identification.

In 2024, Baccichet et al. \cite{Baccichet2024} addressed the challenge of identifying devices with the same model and \gls{os} version by utilizing multi-channel behavior patterns. 
While all these works focus mainly on the accuracy of \gls{pr} fingerprinting, i.e., the capability to link two \gls{pr} to the same device, none of them tackles the problem of compressing such fingerprints.

\section{Methodology}
\label{sec:methodology-fingerprint}
The problem we aim to solve is the following: given a \gls{pr} \( x_n \), we seek to generate a \( M \)-dimensional binary fingerprint \( F(x_n) = [F_1(x_n), F_2(x_n), \ldots, F_M(x_n)] \in \{-1, +1\}^M \), which serves as a robust hash for that \gls{pr} and other \gls{pr} frames from the same physical device.
We adapt the \gls{apb} algorithm for the task at hand. \gls{apb} extends the AdaBoost classifier \cite{Schapire1999} for pairs of samples. Unlike AdaBoost, where each sample in the training set is assigned an individual label, \gls{apb} assigns labels to pairs of samples. These labels indicate whether the pairs are matching or non-matching. The algorithm's goal is to learn a classifier that can determine if a pair of test samples, two \gls{pr} frames in our case, matches or not.

 \begin{algorithm}
     \caption{Asymmetric Pairwise Boosting}
     \label{alg:bamboo}
     \SetKwInOut{Input}{Input}
     \SetKwInOut{Output}{Output}
     \Input{
         \begin{itemize}
             \item Ground truth relationships $\langle x_{a(n)},x_{b(n)}; y_{n} \rangle$
             \item Set of filters $\mathcal{H}=\{h_{1}, h_{2},..., h_{B}\}$
             \item Set of thresholds $\mathcal{T}=\{t_{1}, t_{2},..., t_{T}\}$
         \end{itemize}
     }
     \Output{
         \begin{itemize}
             \item Set of $M<B$ filters $\mathcal{H^{*}}=\{h_{i(1)}, h_{i(2)},..., h_{i(M)}\}$
             \item Corresponding thresholds $\mathcal{T^{*}}=\{t_{j(1)}, t_{j(2)},..., t_{j(M)}\}$
         \end{itemize}
     }
     \SetKwBlock{Beginn}{beginn}{ende}
     \For{$m := 1$ \KwTo $M$} {
         \tcp{Iterate over each bitmask filter}
         \For{$i := 1$ \KwTo $B$} {
             \tcp{Iterate over each threshold}
             \For{$p := 1$ \KwTo $T$} {
                 \tcp{Iterate over each pair}
                     \For{$n := 1$ \KwTo $N$} {
                         $\varepsilon_{i,p} \mathrel{{+}{=}} w_n\delta (h_{i,p}(x_{a(n)},x_{b(n)}) \neq y_n)$
                     }
             }
         }
         \tcp{Select best filter $b_{i(m)}^*$ and best threshold $t_{i(m)}^*$}
         $(i^*, p^*)=\arg\min_{i,p}\varepsilon_{i,p}$
        
         \tcp{Compute the confidence of the weak classifier}
         $c_m \triangleq \log\bigg(\frac{1-\varepsilon^*}{\varepsilon^*}\bigg)$
         \tcp{Asymmetric Weight Update}

         \For{$n := 1$ \KwTo $N$} {
             \If{$y_n=+1$} {
             \If{$h_{i^*,p^*}(x_{a(n)}, x_{b(n)})\neq y_n$} {
                 $w_n \triangleq w_n\cdot \exp{(c_m)}$
             }
             }
         }
         \tcp{Asymmetric Weight Normalization}
         \For{$n := 1$ \KwTo $N$}{
         \If{$y_n = +1$}{ 
             $w_n = \frac{w_n}{\sum_{n : y_n = +1}w_n}$
         }}
         \Return{$(b_{i(m)}^*, t_{i(m)}^*, c_m)$}
     }
 \end{algorithm}

Let \( y_n \in \{-1, +1\} \) denote the label describing the ground truth relationship between the pair \( \langle x_{a(n)}, x_{b(n)} \rangle \), where \( n = 1, \ldots, N \). The label \( y_n \) takes the value \( +1 \) if the two \glspl{pr} are matching (originating from the same device) and \( -1 \) if the frames are from different devices.

Moreover, let \(\mathcal{H}=\{h_{1}, h_{2}, \ldots, h_{B}\}\) define a set of bitmask filters. For each \gls{pr}, we obtain a vector of \(B\) scalar intermediate representations \(\{b_1(x_n), b_2(x_n), \ldots, b_B(x_n)\}\), where each element \(b_i(x_n)\) is the result obtained by filtering the \gls{pr} \(x_n\) with the bitmask filter \(h_i\). %i.e., \(b_i(x_n)=\langle x_n, h_i \rangle\). These intermediate representations are then fed into Algorithm \ref{alg:bamboo} along with the ground truth relationships between each pair of training samples.

The output of the \gls{apb} algorithm consists of a set of \(M<B\) selected filters $F_i$ and corresponding thresholds $t_j$, such that, 

\begin{equation}
\label{eq:descriptor}
F_i(x_n) = 
\begin{cases}
+1 & \text{if } b_{i(m)}(x_n) > t_{j(m)}, \\
-1 & \text{otherwise}.
\end{cases}
\end{equation}

Where $i(m)$ with $m=1, ..., M$, denotes the index of the filter selected during the $m$-th iteration, and $j(m)$ the index of the corresponding threshold in the set $\mathcal{T}=\{t_{1}, t_{2},..., t_{T}\}$.

In the Algorithm \ref{alg:bamboo} intermediate representations are computed as follows:

\begin{equation}
    h_{i,p}(x_{a(n)},x_{b(n)}) \triangleq \xi((b_i(x_{a(n)})-t_p)(b_i(x_{b(n)})-t_p))
\end{equation}

Where $\xi$ is defined as:

\begin{equation}
\label{xi-function}
\xi((b(x_{a})-t)(b(x_{b})-t)) = 
\begin{cases}
+1 & \text{if } b(x_{a,b}) > t, \\
-1 & \text{otherwise}.
\end{cases}
\end{equation}

Each step of the boosting algorithm selects the filter and the threshold that provide the best possible separation between matching and non-matching \glspl{pr}. This involves choosing the weak classifier that minimizes the weighted error function over the set of all training \glspl{pr}. The assignment of a weight to each pair of \glspl{pr} is intended to emphasize the importance of misclassified pairs. Specifically, the more accurately the previously selected weak classifiers can classify a pair \(\langle x_{a(n)},x_{b(n)}\rangle\), the lower the corresponding weighting factor \(w_n\) becomes. %Notably, only the weights corresponding to matching pairs are updated throughout the execution of the \gls{apb} algorithm, as described in Algorithm \ref{alg:bamboo}. As demonstrated by \cite{Sukthankar2006}, a symmetric weighting of matching and non-matching training \glspl{prb} would lead to the violation of Adaboost’s weak classifier criterion, resulting in lower performance.

\subsection{Candidate bitmask filters}
\label{methodology-fingerprint:filters}

The set of candidate filters contains filters of four different types (A,B,C and D). Each filter has length \(L\), preceded by a prefix of \(P\) zeros and followed by a suffix of \(S\) zeros, such that $P+L+S=1784$ bits and matches the size of a \gls{pr}. Type A filters are characterized by the first half of non-zero bits being $+1$, and the second half being $-1$. Type B filters are the complement of Type A, with the first half being $-1$ and the second half $+1$. In Type C filters, all non-zero bits are $-1$, while in Type D, all non-zero bits are $+1$. Our objective was to replicate Haar-like features, generally used in image processing, using bitmasks. Similar to Haar-like features, our filters can be defined as the difference of the sum of the bits within specific areas of a \gls{pr} covered by the bitmask. These areas can be located at any position within the \gls{pr} and have a length \(L \in \{4, 8, 16\}\). We used a total of 2,492 different filters, varying the position of the $L$ filtering bits throughout the complete length of a \gls{pr}. These filters were designed to select entire bytes, portions of fields, or overlapping portions of data across different fields of a \gls{pr}.

%Figure \ref{fig:filter-types} describes the type of filters we used. These filters are bitmasks of length \(L\), preceded by a prefix of \(P\) zeros and followed by a suffix of \(S\) zeros. Our objective was to replicate Haar-like features, generally used in image processing, using bitmasks. 

%Type A filters are characterized by the first half of non-zero bits being $+1$, and the second half being $-1$. Type B filters are the complement of Type A, with the first half being $-1$ and the second half $+1$. In Type C filters, all non-zero bits are $-1$, while in Type D, all non-zero bits are $+1$.

%Similar to Haar-like features, our filters can be defined as the difference of the sum of the bits within specific areas of a \gls{prb} covered by the bitmask. These areas can be located at any position within the \gls{prb} and have a scale \(L \in \{4, 8, 16\}\). Each feature type can indicate the presence or absence of certain characteristics, which can be more relevant than others to discriminate between \glspl{prb}, hence creating an efficient and compact fingerprint.

\subsection{Fingerprint Matching}
\label{subsec:matching}

The \gls{apb} algorithm assigns a weight, namely \textit{confidence}, $c_m$ with $m=1, ..., M$ to each of the selected weak classifiers. Such weights are subsequently exploited to compute a weighted Hamming distance between a pair of test \glspl{pr}, $x_{n_1}$ and $x_{n_2}$, according to

\begin{equation}
\label{eq:weighted-h-distance}
    H(x_{n_1}, x_{n_2}) = \sum_{m=1}^{M}c_m \xi_m(x_{n_1}, x_{n_2})
\end{equation}

Where the $\xi$ function is described in \eqref{xi-function} and assumes value $+1$ when both filtered \glspl{pr} fall on the same side of the threshold corresponding to the selected filter, $-1$ otherwise.

Then $H(x_{n_1}, x_{n_2})$ is compared to a threshold $\tau$ to determine if $\langle x_{n_1}, x_{n_2} \rangle$ is a matching pair. The value of $\tau$ is selected based on the desired trade-off between \gls{tpr} and \gls{fpr}.

We note that ignoring these weights by setting $c_m =1$ does not significantly impact the discriminative power. At the same time, this change allows to use the simple Hamming distance for matching two fingerprints, which can be efficiently performed in hardware.

%The fast computation of Hamming distances, thanks to optimized instruction sets and XOR gates, is key to the matching process and it is one of the advantages of a binary fingerprint. In Section \ref{sec:experimental-results} we we will show that ignoring these weights by setting $c_m =1$ does not significantly impact the discriminative power, while enabling the use of simple Hamming distance.

For that reason, we can simplify the prediction process: two \glspl{pr} having fingerprints $F(x_1)$ and $F(x_2)$ are predicted as matching ($+1$) or non-matching ($-1$) according to the following equation:

\begin{equation}
\hat{y} = 
\begin{cases}
+1 & \text{if } \sum F(x_1) \oplus F(x_2) < \tau, \\
-1 & \text{otherwise}.
\end{cases}
\end{equation}

\begin{figure}[t]
   \centering
   \includegraphics[width=0.85\linewidth]{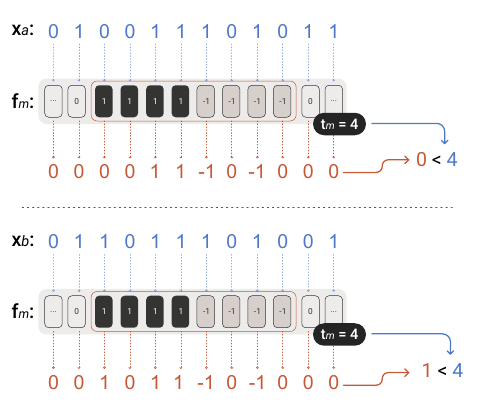}
   \caption{Filtering process for a matching pair of \glspl{pr}. The prediction is also matching: both of the filtered \glspl{pr} fall on the same side of the threshold.}
   \label{fig:prediction-process}
\end{figure}

Where $\hat{y}$ is the prediction (matching / non-matching pair) and $\tau\in [0,M]$. Figure \ref{fig:prediction-process} illustrates the matching process for one bit of the
fingerprint.

%Figure \ref{fig:prediction-process} illustrates the matching process for one bit of the fingerprint. First, we perform the bitwise multiplication of the first \gls{prb} \( x_a \) with the selected filter. Simultaneously, we perform the same operation between the second \gls{prb} \( x_b \) and the same filter. We then compare the results to the threshold associated with the selected filter. If both sums are on the same side of the threshold (either both lower or both above the threshold) for that particular filter, the \glspl{prb} are considered matching.

\section{Dataset}
\label{sec:dataset}

% \begin{table}[t!]
% \caption{List of selected Information Elements and relative lengths.}
% \label{tab:ie-binary-dissected}
% \begin{center}
% \begin{tabular}{|c|c|c|}
% \hline
% \textbf{Element} & \textbf{Element ID}& \textbf{Length (\textit{bits})} \\
% \hline
% \hline
% Lenght HT Capabilities & 45 & 8 \\
% \hline
% HT Capabilities & 45 & 208 \\
% \hline
% Length Extended Capabilities & 127 & 8 \\
% \hline
% Extended Capabilities & 127 & 88 \\
% \hline
% Length Vendor Specific Tags & 221 & 8 \\
% \hline
% Vendor Specific Tags & 221 & 1336 \\
% \hline
% DSSS Parameter (concat.) & 3 & 128 \\
% \hline
% \end{tabular}
% \end{center}
% \end{table}
 
We use two open datasets from previous studies, specifically \cite{Pintor2022_Dataset} and \cite{Baccichet2024}, to ensure the reproducibility of our results. The two starting datasets are merged in a single one which includes raw \glspl{pr} stored in \texttt{.PCAP} file format, and contain labeled information from 33 different devices (i.e., each \gls{pr} is labeled with the originating device identifier). Some of these devices share the same model and \gls{os} version, simulating a realistic scenario where identical devices need to be distinguished as separate entities. The data were collected on three non-overlapping channels in the 2.4 GHz Wi-Fi band, specifically channels 1, 6, and 11.  Following the IEEE 802.11 standard, we developed a custom packet dissector to extract fields from the \glspl{pr} in binary format. Details on the device contained in the dataset are reported in Table \ref{tab:dataset-devices}.

\begin{table}[t!]
\caption{Dataset's Devices Details}
\label{tab:dataset-devices}
\begin{center}
\begin{tabular}{|l|l|c|}
\hline
\textbf{Device Model} & \textbf{\gls{os} Version} & \textbf{No. Entries} \\
\hline
\hline
Apple iPhone 6 & iOS \texttt{12.05.02} & 1 \\
\hline
Apple iPhone 7 & iOS \texttt{14.6, 15.5} & 2 \\
\hline
Apple iPhone 11 & iOS \texttt{16.4.1(a)} & 4 \\
\hline
Apple iPhone 12 & iOS \texttt{14.05, 16.4.1(a)} & 3 \\
\hline
Apple iPhone XR & iOS \texttt{14.6, 16.4.1(a)} & 3 \\
\hline
Apple iPhone XS & iOS \texttt{14.05} & 1 \\
\hline
\hline
Google Pixel 3A & Android \texttt{11} & 2 \\
\hline
Huawei L21 & Android \texttt{6.0} & 1 \\
\hline
Huawei Honor 9 & Android \texttt{9} & 1 \\
\hline
Huawei P10 & Android \texttt{9} & 1 \\
\hline
Huawei P20 & Android \texttt{10} & 1 \\
\hline
OnePlus Nord & Android \texttt{11} & 1 \\
\hline
Oppo Find X3 Neo & Android \texttt{13} & 1 \\
\hline
Samsung Galaxy J6 & Android \texttt{10} & 1 \\
\hline
Samsung Galaxy M31 & Android \texttt{11} & 1 \\
\hline
Samsung Galaxy S4 & Android \texttt{4.02.02} & 1 \\
\hline
Samsung Galaxy S6 & Android \texttt{7} & 1 \\
\hline
Samsung Galaxy S7 & Android \texttt{8} & 1 \\
\hline
Samsung Galaxy S21 & Android \texttt{13} & 1 \\
\hline
Xiaomi Mi A2 Lite & Android \texttt{10} & 1 \\
\hline
Xiaomi Redmi 4 & Android \texttt{6.00.01} & 1 \\
\hline
Xiaomi Redmi 5 Plus & Android \texttt{8.01} & 1 \\
\hline
Xiaomi Redmi Note 7 & Android \texttt{10} & 1 \\
\hline
Xiaomi Redmi Note 9S & Android \texttt{11} & 1 \\
\hline
\end{tabular}
\end{center}
\end{table}

% \subsection{Binary Dissection}
% \label{dataset:dissection}

% Following the IEEE 802.11 standard, we developed a custom packet dissector to extract fields from the \glspl{pr} in binary format. This dissector, implemented using Python's Scapy library, processes each \gls{pr} in the dataset by identifying the \gls{ie} IDs specified by the standard, along with fixed and variable-length fields. The dissector extracts these fields in binary format for the entire packet length and stores them in a Comma Separated Values file (CSV), with each field reported in a separate column. This process facilitates subsequent data analysis and ensures that all relevant information is retained in a well-structured format.

%\begin{figure}[t]
%    \centering
%    \includegraphics[width=1\linewidth]{figures/Flow Diagram.pdf}
%    \caption{Fingerprint Extraction Pipeline.}
%    \label{fig:fingerprint-generation-pipeline}
%\end{figure}

\subsection{Data Pre-Processing}
\label{dataset:pre-processing}

We pre-process the binary \glspl{pr} by considering only the \glspl{ie} which have been demonstrated to be relevant by the state of the art, particularly following the insights from \cite{Pintor2022}, \cite{Baccichet2024}, \cite{Fenske2021}, and \cite{Uras2020}. The most relevant fields for the defined use case are \textit{HT Capabilities}, \textit{Extended Capabilities}, and \textit{Vendor Specific Tags}. The first two fields are used by the transmitting device to advertise its technical capabilities, while the latter carries additional information not specified by the standard.

%Since it has been proven that other fields are less useful for creating a fingerprint, and because our goal is to develop a compact and efficient fingerprint, we decided to retain only the fields listed in Table \ref{tab:ie-binary-dissected}, reported with the corresponding length in bits.

%To incorporate the findings from \cite{Baccichet2024} on the multi-channel behavior of devices, we processed the data as follows. First, we grouped the data by \gls{mac} address to move from individual \gls{pr} to burst of \gls{pr} sharing the same content. This approach is feasible because the \glspl{ie} remain stable within a single \gls{prb}, allowing for consistent data aggregation. Additionally, since a burst of probes is sent in less than a second, we can buffer the \gls{mac} address, group by it, and then discard it.

%We complete the set of \glspl{ie} considered by concatenating the DSSS Parameter Set's (Element ID 3) \textit{Current Channel} field, referred to as DS Channel, from the various \glspl{pr} within the burst, with a maximum length of 128 bits per burst. The DSSS Parameter Set provides channel number identification for stations in compliance with the IEEE 802.11 standard, and its \textit{Current Channel} field is 8 bits long. This process includes the DS Channel field from the first $n \leq 16$ \glspl{pr} within each burst. Each field is then zero-padded to the right up to the maximum length to ensure uniform column alignment.

Finally, we concatenated all the columns to generate the raw binary data to be processed by our algorithm using specific filters. After the pre-processing step, the size of each \gls{pr} is equal to 223 bytes (1784 bits).

\section{Experimental Results}
\label{sec:experimental-results}

%Our analysis has two main goals: (i) investigating the potential of the \gls{apb} algorithm to create a compact fingerprint for \glspl{pr}. In parallel, the second goal involves assessing the performance of state-of-the-art methods in the context of \gls{mac} address de-randomization, with the aim of developing a storage-accuracy framework to evaluate our methodology. By doing so, we aim to provide valuable insights into the comparative effectiveness of existing techniques, also when dealing with devices that share the same model and \gls{os}. This analysis contributes to the broader understanding of Wi-Fi device tracking and identification in real-life scenarios.

%\subsection{Experimental Setup}
%\label{experimental-results:experimental-setup}

Our experiments are articulated in two steps: first, we identify the most interesting portions of a \gls{pr} to create a compact fingerprint (i.e., selecting $M$ filters), using the \gls{apb} algorithm. This involves creating a training dataset of pairs of \glspl{pr}, a testing dataset, as well as investigating the threshold $\tau$ to balance \gls{fpr} and \gls{tpr}.
In the second phase, we assess the performance of our newly-developed binary fingerprint against state-of-the-art techniques that have proven efficient at discriminating \glspl{pr}, in tasks such as counting devices in a room via clustering of \glspl{pr} fingerprints. In particular, we compare our methodology with the techniques described in \cite{Pintor2022} and \cite{Baccichet2024}.

%To test the performance of the proposed methods at different sizes of the device population, we adopt the following strategy: for each target population size $p = 1, \dots, P-1$, where $P$ is the maximum number of devices in the original dataset, we produce $d=10$ different subsets by selecting $p$ devices at random. We then run three different algorithms (the one described in \cite{Baccichet2024}, the one described in \cite{Pintor2022}, and the one proposed in this work) over each subset, keeping track of the final Homogeneity, Completeness, and V-Measure as well as the \gls{rmse} between the final number of created clusters and $p$. Since the works in \cite{Baccichet2024} and \cite{Pintor2022} require hyperparameter tuning, we use the best parameters declared in the respective papers.

\begin{table}[t!]
\caption{Top 16 best filter-threshold configurations and relative Information Elements Ordered by Confidence}
\label{tab:best-configs}
\begin{center}
\begin{tabular}{|c|c|c|c|c|c|c|}
\hline
\textbf{Type} & \textbf{T} & \textbf{L} & \textbf{P} & \textbf{S} &  \textbf{Information Elements} \\
\hline
\hline
D & 5 & 16 & 336 & 1432 & Vendor Specific Tags \\
\hline
B & 2 & 16 & 16 & 1752 & HT Capabilities \\
\hline
D & 3 & 8 & 24 & 1752 & HT Capabilities \\
\hline
D & 4 & 16 & 16 & 1752 & HT Capabilities \\
\hline
D & 3 & 8 & 336 & 1440 & Len. Vendor Specific Tags \\
\hline
B & 8 & 8 & 8 & 1768 & HT Capabilities \\
\hline
D & 1 & 8 & 328 & 1448 & Vendor Specific Tags \\
\hline
D & 1 & 16 & 352 & 1416 & Vendor Specific Tags \\
\hline
A & 3 & 16 & 336 & 1432 & Vendor Specific Tags \\
\hline
D & 2 & 16 & 240 & 1528 & Extended Capabilities \\
\hline
B & 1 & 8 & 216 & 1560 & Len. Extended Capabilities \\
\hline
B & 1 & 16 & 0 & 1768 & HT Capabilities* \\
\hline
D & 4 & 8 & 8 & 1768 & HT Capabilities \\
\hline
D & 1 & 8 & 248 & 1528 & Extended Capabilities \\
\hline
D & 1 & 16 & 368 & 1400 & Vendor Specific Tags \\
\hline
D & 7 & 16 & 0 & 1768 & HT Capabilities* \\
\hline
\end{tabular}
\\~\\
\footnotesize{* Length HT Capabilities + First 8 bits of HT Capabilities}\\
\end{center}
\end{table}

\subsection{Training Data}
\label{experimental-results:training-data}

Since the algorithm we use considers pairs of \glspl{pr} and classifies them as matching or non-matching, we require both a training set and a test set containing pairs of \glspl{pr}, where each pair is either matching or non-matching. To achieve this, we performed the following:

\begin{itemize}
    \item \textbf{Matching Pairs Set}: leveraging the labeled dataset, we identified 1000 unique pairs of \glspl{pr} from the same devices to serve as matching examples.
    \item  \textbf{Non-Matching Pairs Set}: We created non-matching pairs by selecting \glspl{pr} from different devices. This ensured that the pairs represented distinct devices, facilitating the algorithm's ability to differentiate between them.
    \item \textbf{Balancing Dataset}: We ensured that both matching and non-matching pairs were evenly represented to avoid bias and improve the algorithm's generalization capability.
    \item \textbf{Dataset Partitioning}: The dataset was divided into training and test subsets (60-40) maintaining the balance of matching and non-matching pairs in each subset. The training dataset is used as input to the \gls{apb} algorithm, while the test set is used to compute to evaluate performance and possibly tune matching hyperparameters.
\end{itemize}

\subsection{Selected Filters}
\label{experimental-results:selected-filters}

The filters selected by the \gls{apb} algorithm are shown in Table \ref{tab:best-configs} when \( M = 16 \), as an example. From 2,492 candidate filters, 16 filters and their associated thresholds are selected. As shown in Table \ref{tab:best-configs}, these 16 filters vary in type and size. The table provides details on the filter type, relative threshold \( T \), length \( L \), zero prefix length \( P \), zero suffix length \( S \), and the \gls{ie} considered by the filter. 

From this analysis, it is evident that the most relevant and discriminating portion of the \gls{pr} resides in the \textit{Vendor Specific Tags} (Element ID \texttt{221}). The filters are listed in the table, ordered by descending confidence \( c_m \). There is minimal redundancy among the selected filters. Filters with longer lengths (16 or 8 bits) are predominantly chosen. %It is interesting to notice that the majority of the selected filters (22 out of 32) are of the D-type, which means that we are considering non other than the Hamming distance of the considered slices of the \gls{pr}.

\subsection{Descriptor Length}
\label{experimental-results:descriptor-length}

\begin{figure}[t]
    \centering
    \includegraphics[width=1\linewidth]{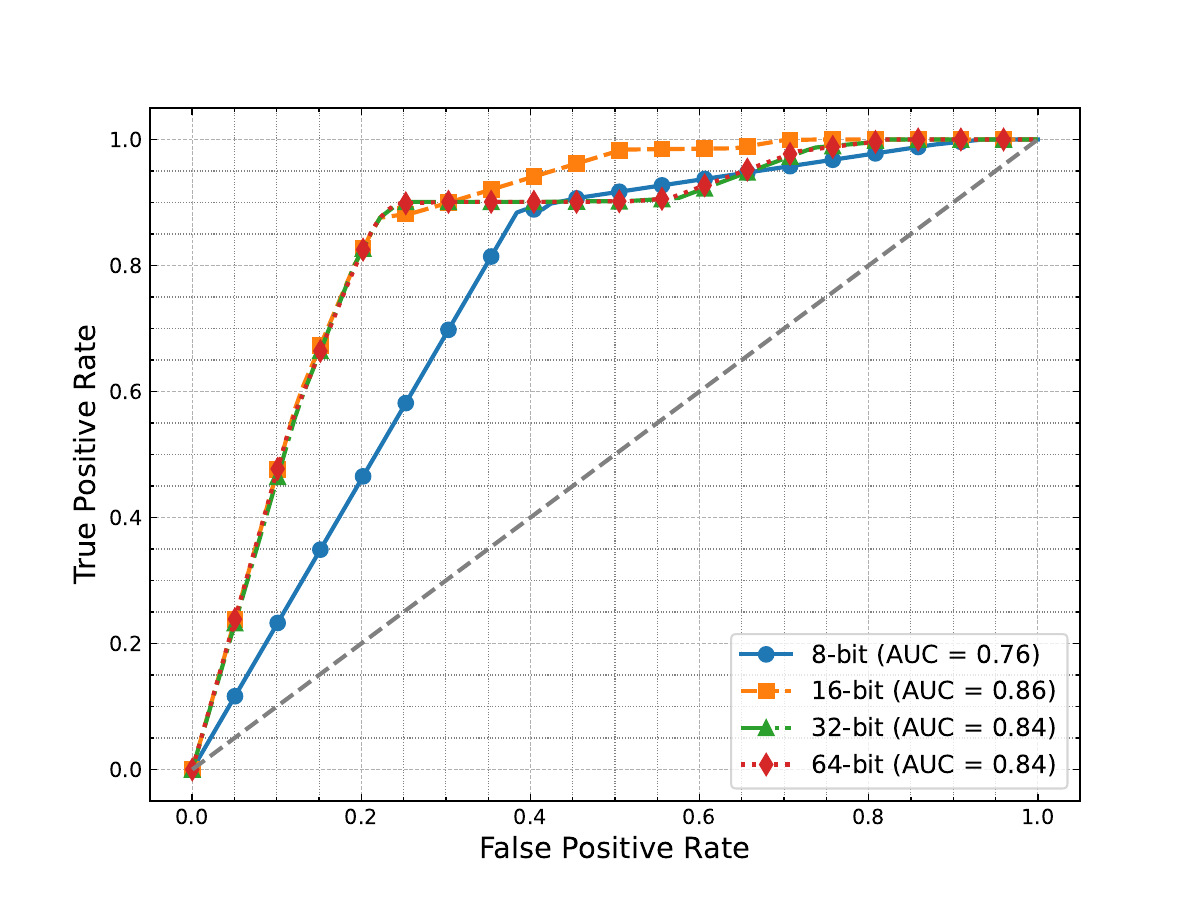}
    \caption{ROC w/o Weighted Hamming Distance}
    \label{fig:ROC}
\end{figure}

We investigated the impact of the size of the descriptor on the discriminative power by varying \( M = 8, 16, 32, 64 \) bits. Figure \ref{fig:ROC} shows the results in terms of \glspl{roc} curves, using the test dataset and varying the threshold \(\tau\) described in \ref{subsec:matching} from 0 to \( M \). We then identified the optimal \(\tau\) for each specific fingerprint length as the value corresponding to the point on the ROC curve closer to the top-left corner, that is the best trade-off between TPR and FPR.

From Figure \ref{fig:ROC}, it can be observed that the 16-bit fingerprint slightly outperforms the other tested alternatives. We explain this behavior by noting that using a larger set of filters, and consequently a longer fingerprint, introduces more ambiguity into the fingerprint, which negatively impacts the ability to accurately predict the (non) matching status of \gls{pr} pairs.

% \subsection{Weighted Hamming Distance}
% \label{experimental-results:weighted-hamming-distance}

% % \begin{figure}[t]
% %     \centering
% %     \includegraphics[width=1\linewidth]{figures/ROC-comparison_h.pdf}
% %     \caption{ROC w/ Weighted Hamming Distance}
% %     \label{fig:ROC-h-distance}
% % \end{figure}

% As mentioned in Section \ref{subsec:matching}, the \gls{apb} algorithm assigns a weight, namely \textit{confidence}, \( c_m \), to each selected filter. This weight can be used to compute a weighted version of the Hamming distance, as indicated in Equation \ref{eq:weighted-h-distance}. Although the computation of the weighted Hamming distance can be optimized, as recently discussed in \cite{Fan2013}, the unweighted version is still preferable when matching is performed in large-scale databases. Therefore, we also investigated how the performance varies when neglecting the weights when matching descriptors. 

% We observed that the two approaches achieve very similar performance, with an unexpected slight improvement noted when not using the weighted version. For completeness, we reported the best \(\tau\) values in Table \ref{tab:best-taus} also when not using the weighted Hamming distance, by setting \( c_m = 1 \). However, since the performance was slightly better with the non-weighted version, we proceeded with it in the following experiments.

% \begin{table}[t!]
% \caption{List of $\tau$ maximizing \gls{auc} in the \gls{roc} plot}
% \label{tab:best-taus}
% \begin{center}
% \begin{tabular}{|c|c|c|}
% \hline
% \textbf{Bits} & $\boldsymbol{\tau}$ \textbf{w/ Hamming Distance} & $\boldsymbol{\tau}$ \textbf{w/o Hamming Distance} \\
% \hline
% \hline
% 8 & 1.00 & 1.00 \\
% \hline
% 16 & 1.94 & 2.00 \\
% \hline
% 32 & 7.32 & 7.00 \\
% \hline
% 64 & 6.70 & 8.00 \\
% \hline
% \end{tabular}
% \end{center}
% \end{table}

\subsection{Clustering Results}
\label{experimental-results:comparative-clustering}

As stated in the introduction, our secondary goal, other than obtaining a compact and fast descriptor for Wi-Fi \glspl{pr}, was to compare its discriminative performance against state-of-the-art techniques for device counting. Such techniques exploit traditional clustering algorithms, such as \gls{dbscan}, applied to \glspl{ie} \cite{Uras2020}, \cite{Pintor2022} or channel behavior \cite{Baccichet2024} fingerprints.

\begin{figure}[t]
    \centering
    \includegraphics[width=1\linewidth]{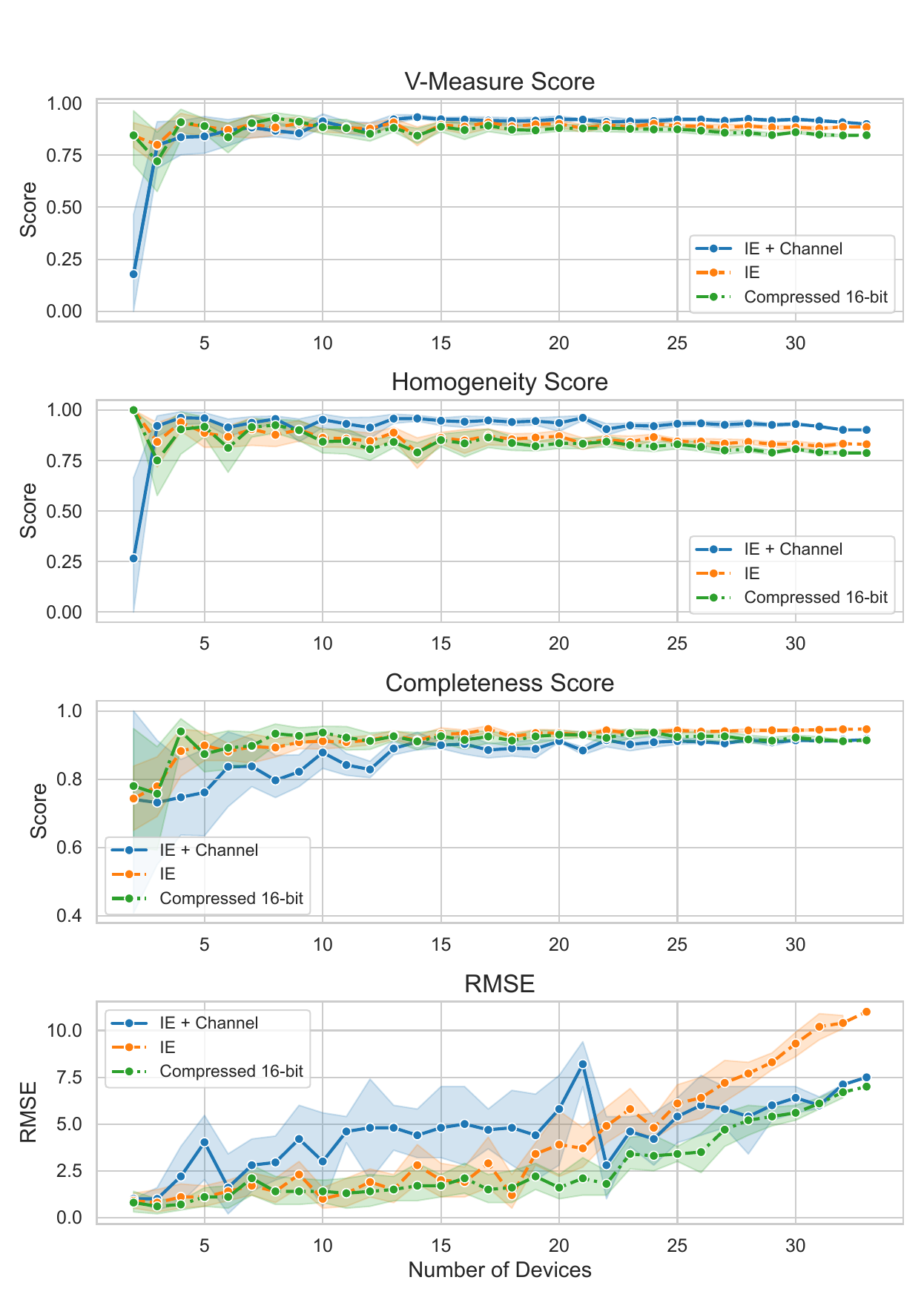}
    \caption{Clustering Metrics of the pairwise boosting clustering}
    \label{fig:clustering-metrics-sota}
\end{figure}

\begin{table*}[t]
\centering
\caption{Table of Best Filters, Best Thresholds, Minimum Errors, and Confidence}
\label{tab:clustering-metrics-comparison}
\begin{tabular}{|l||c|c|c|c|c|}
\hline
\textbf{Metric} & \textbf{IE + Channel} & \textbf{IE} & \textbf{Fingerprint 16-bit} & \textbf{Fingerprint 32-bit} & \textbf{Fingerprint 64-bit} \\
\hline
\hline
\textbf{V-Measure} (avg) & 0.875 & 0.884 & 0.869 & 0.868 & 0.841 \\
\hline
\textbf{Homogeneity} (avg) & 0.913 & 0.862 & 0.840 & 0.827 & 0.842 \\
\hline
\textbf{Completeness} (avg) & 0.866 & 0.914 & 0.910 & \textbf{0.928} & 0.852 \\
\hline
\textbf{RMSE} (avg) & 4.446 & 3.871 & \textbf{2.543} & \textbf{2.299} & \textbf{3.202} \\
\hline
\textbf{Memory} (bits / probe request) & 1784 & 1656 & \textbf{16} & \textbf{32} & \textbf{64} \\
\hline
\hline
\textbf{\textit{Compression Ratio}} & 1 & 1.07 & \textbf{111.5} & \textbf{55.7} & \textbf{27.8} \\
\hline
\end{tabular}
\end{table*}

To test the performance of the proposed methods at different sizes of the device population, we adopt the following strategy: for each target population size $p = 1, \dots, P-1$, where $P$ is the maximum number of devices in the original dataset, we produce $d=10$ different subsets by selecting $p$ devices at random. We then run three different clustering algorithms (the one described in \cite{Baccichet2024}, the one described in \cite{Pintor2022}, and the one proposed in this work) over each subset, keeping track of the final cluster Homogeneity, Completeness, and V-Measure as well as the \gls{rmse} between the final number of created clusters and $p$. 

For what concerns the proposed method, clustering is performed in an online fashion: the first \gls{pr} of the dataset is assigned to the first cluster and the corresponding fingerprint is stored as cluster identifier. For the following \gls{pr}s, the computed fingerprint is compared with the existing clusters fingerprint using the Hamming distance. A new cluster is created if no existing cluster has Hamming distance smaller than $\tau$.
As expected from the evaluation done in Section \ref{experimental-results:descriptor-length}, the shortest fingerprint performed the worst, while the other three lengths performed comparably, with the 16-bit length being the best. To be comparable with state-of-the-art techniques, we calculated clustering metrics, such as Homogeneity, Completeness, V-Measure, and \gls{rmse}. Results for the 16, 32, and 64-bit long fingerprints are reported in Table \ref{tab:clustering-metrics-comparison}; we omit the 8-bit long one, as its metrics were the worst and thus not considered for the rest of the experiments.

For what concerns the methods in \cite{Pintor2022} and \cite{Baccichet2024}, labeled as \textit{IE} and \textit{IE + Channel}, respectively, in Figure \ref{fig:clustering-metrics-sota}, they work by applying DBSCAN clustering to the raw \gls{pr} \glspl{ie}. Since these methods require hyperparameters tuning for DBSCAN, we use the best parameters declared in the respective papers.
The size of the input to such algorithms are the raw \gls{pr} data, including channel information (1784 bits per \gls{pr}) for the method in \cite{Baccichet2024}, while slightly less (1656 bits per \gls{pr} for the one in \cite{Pintor2022}.

%We compared our proposed fingerprinting technique, specifically the 16-bit version, with the state-of-the-art methods described in \cite{Pintor2022} and \cite{Baccichet2024}. These methods are labeled as \textit{IE} and \textit{IE + Channel}, respectively, in Figure \ref{fig:clustering-metrics-sota}. 

% \begin{figure}[t]
%     \centering
%     \includegraphics[width=1\linewidth]{figures/clustering_metrics_comparison.pdf}
%     \caption{Clustering Metrics using different lengths of the fingerprints obtained using the pairwise boosting algorithm.}
%     \label{fig:clustering-metrics-length}
% \end{figure}

The detailed results are presented in Figure \ref{fig:clustering-metrics-sota} and are further summarized in Table \ref{tab:clustering-metrics-comparison}, where we provide averaged values for a comprehensive comparison. Our approach, while slightly lagging behind the state-of-the-art methods in terms of V-Measure, demonstrates a significant improvement in the accurate identification of the number of distinct devices within the dataset, showing the lowest RMSE. This indicates that our method may offer superior granularity or sensitivity in distinguishing between devices, which can be crucial in applications requiring precise device counts.

%\subsection{Discussion}
%\label{experimental-results:discussion}

A particularly noteworthy advantage of our proposed method is its remarkable efficiency in memory usage. Our fingerprinting technique requires only about $1\%$ of the memory compared to the techniques described in \cite{Pintor2022} and \cite{Baccichet2024}. 

% This substantial reduction in memory consumption can lead to more scalable implementations, especially in environments with constrained resources, such as embedded systems or large-scale deployments.

The implications of these findings suggest that while there may be a slight trade-off in identification accuracy (lowered by about $1,6\%$ in the V-Measure), the benefits of reduced memory usage and improved device count accuracy can make our method particularly valuable in specific contexts. For instance, scenarios where memory is a critical constraint or where accurate device enumeration is prioritized over individual device identification could particularly benefit from our approach. Further research could explore optimizing the trade-offs between memory usage, identification accuracy, and other relevant performance metrics to enhance the applicability of our technique across different use cases.

\section{Conclusion}
\label{sec:conclusion}
\gls{pr} fingerprinting enables the defeat of MAC randomization, facilitating applications such as device counting, occupancy estimation, and urban flow monitoring. This paper proposes a compact yet discriminative \gls{pr} fingerprint, whose structure is learned through a ML technique called \gls{apb}. By leveraging a dataset of \gls{pr} frames from various devices (including some of identical types), the proposed methodology identifies the discriminative parts of a \gls{pr} and compresses them into a single bit using specifically learned thresholds.

The resulting binary fingerprint, whose size is fully customizable by the user, significantly reduces memory requirements for processing \gls{pr} frames without compromising fingerprinting accuracy. Experiments conducted on publicly available datasets demonstrate that a fingerprint as short as 16 bits can achieve performance comparable to state-of-the-art solutions while using 100 times less memory space. To support reproducible research, the source code of our methodology is made publicly available.

\balance

\section*{Acknowledgement}
This study was carried out within the PRIN project COMPACT and received funding from Next Generation EU, Mission 4 Component 1, CUP: D53D23001340006.

\bibliographystyle{IEEEtran}
\bibliography{IEEEabrv.bib, bibliography.bib}

\end{document}